\documentclass[aps,prl,twocolumn,amsmath,superscriptaddress,reprint]{revtex4-2}
\usepackage{CJK}
\usepackage{amssymb,amsmath}
\usepackage{graphicx,color}
\usepackage{lipsum}
\usepackage{ulem}
\usepackage{hyperref}
\usepackage{lineno}

\usepackage{ulem} 
\usepackage{soul}

\begin{document}

\begin{CJK*}{UTF8}{gbsn}
 
\title{Jamming as a topological satisfiability transition with contact number hyperuniformity and criticality} 

\author{Jin Shang}
    \affiliation{School of Physics and Astronomy, Shanghai Jiao Tong University, 800 Dong Chuan Road, 200240 Shanghai, China.}
\author{Yinqiao Wang}
    \affiliation{Research Center for Advanced Science and Technology, University of Tokyo, 4-6-1 Komaba, Meguro-ku, Tokyo 153-8505, Japan.}
\author{Deng Pan}
    \affiliation{Institute of Theoretical Physics, Chinese Academy of Sciences, Beijing 100190, China. }
\author{Yuliang Jin}
    \email[Email address: ]{yuliangjin@mail.itp.ac.cn}
    \affiliation{Institute of Theoretical Physics, Chinese Academy of Sciences, Beijing 100190, China. }
    \affiliation{School of Physical Sciences, University of Chinese Academy of Sciences, Beijing 100049, China.}
    \affiliation{Wenzhou Institute, University of Chinese Academy of Sciences, Wenzhou 325000, China.}
\author{Jie Zhang}
    \email[Email address: ]{jiezhang2012@sjtu.edu.cn}
    \affiliation{School of Physics and Astronomy, Shanghai Jiao Tong University, 800 Dong Chuan Road, 200240 Shanghai, China.}
    \affiliation{Institute of Natural Sciences, Shanghai Jiao Tong University, 200240 Shanghai, China.}
    
\begin{abstract}

The jamming transition between flow and amorphous-solid states exhibits paradoxical properties characterized by hyperuniformity (suppressed spatial fluctuations) and criticality (hyperfluctuations), whose origin remains unclear. Here we model the jamming transition by a topological satisfiability transition in a minimum network model with simultaneously  hyperuniform distributions of contacts, diverging length scales and scale-free clusters.
We show that these phenomena stem from isostaticity and mechanical stability: the former imposes a global equality, and the latter local inequalities on arbitrary sub-systems. This dual constraint bounds contact number fluctuations from both above and below, limiting them to scale with the surface area.  The hyperuniform and critical exponents of the network model align with those of frictionless jamming, suggesting a new universality class of non-equilibrium phase transitions. Our results provide a minimal, dynamics-independent framework for jamming criticality and hyperuniformity in disordered systems.

\end{abstract}
\maketitle
\end{CJK*}


\begin{figure*}[t]
    \centering
    \includegraphics[width= 17.8 cm]{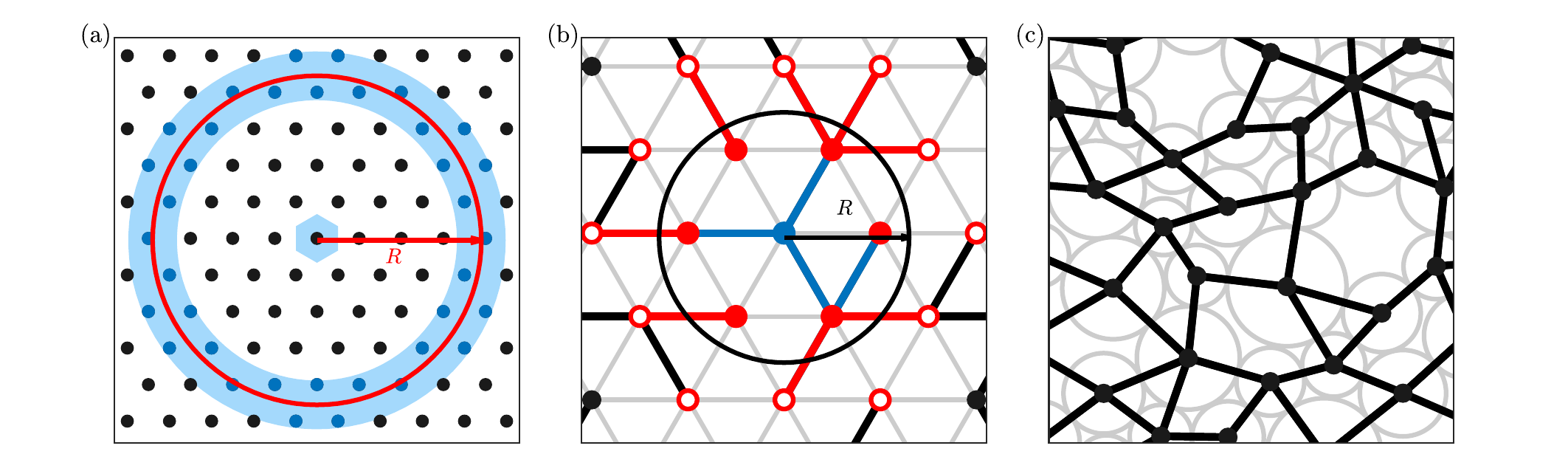}
    \caption{
    {\bf Schematic diagrams of  hyperuniform systems.}
(a) A schematic 2D crystal  illustrating DH: fluctuations of particle numbers  within an observation window of radius $R$ primarily arise from particles in the boundary region (blue ring), 
scaling as $\sigma_n^2(R) \propto R^{d-1}$. 
(b) Net-I  on a triangular lattice. Dots are sites; thin gray edges indicate all possible connections, and thicker edges form a Net-I configuration. The black circle denotes an observation window of radius $R$; solid dots are internal sites. Here, $n = 7$ (solid red and blue dots), $n_{\rm E} = 6$ (solid red), $n_{\rm O} = 12$ (hollow red), $m_{\rm I} = 4$ (blue edges), $m_{\rm B} = 8$ (red edges), and the total contact number is $z(s) = 16$. 
(c) Contact network in a jammed packing of frictionless particles.
    }
    \label{fig:diagram}
\end{figure*}

\begin{figure*}[!t]
    \centering
    \includegraphics[width= 17.8 cm]{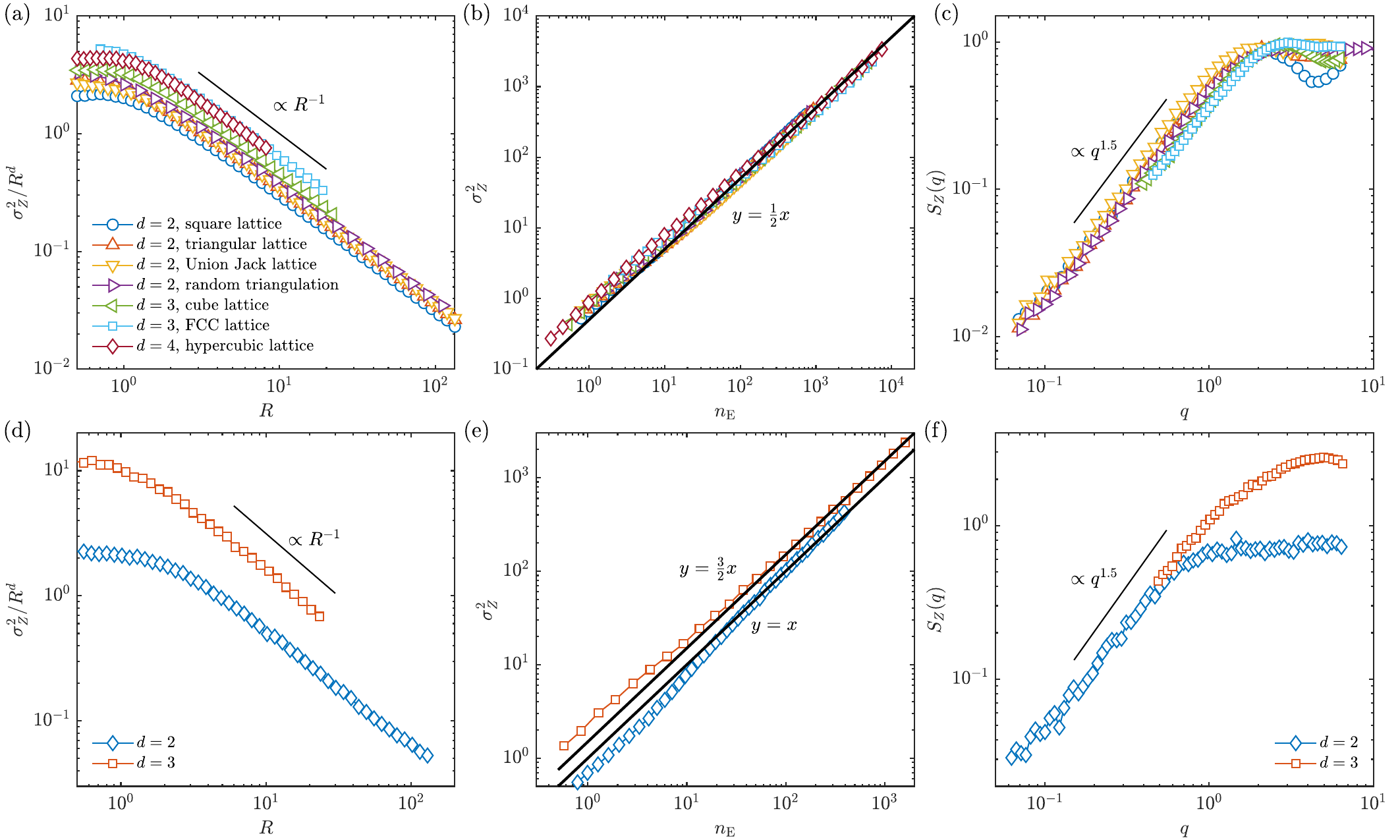}
    \caption{
    {\bf CNH in (a-c) Net-I and (d-f) packings.} 
    (a,d) The variance $\sigma_Z^2$, normalized by $R^d$, versus radius $R$ for various site patterns and dimensionality $d$, showing a universal scaling $\sigma_Z^2 \propto R^{d-1}$ at large scales.
    (b,e) $\sigma_Z^2$ versus the number of boundary sites $n_{\rm E}$. Solid black lines represent the indicated linear functions.
    (c,f) Contact-number fluctuations in the Fourier space, characterized by the structure factor $S_Z(q)$.
    }
    \label{fig:fluactuation}
\end{figure*}

\begin{figure}[tb]
    \centering 
    \includegraphics[width= 8.6 cm]{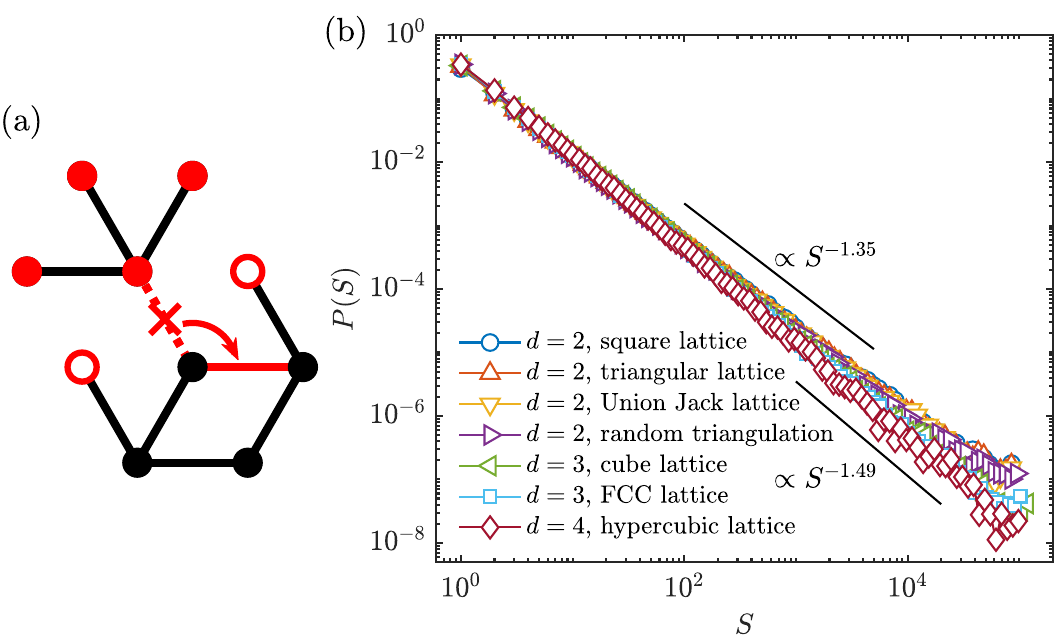}
    \caption{{\bf Unstable clusters upon perturbations at the TST in Net-I.}
    (a) A schematic diagram illustrating the perturbation realized by swapping a bond.
    Prior to breaking the marked edge (`X'), the network satisfies the global constraint $ M = N - 1 $, with $ M = 9 $ and $ N = 10 $. After removing the marked edge, a four-node cluster (solid red circles), denoted by $ s_0 $, becomes isolated and violates the local stability condition $ m(s_0) \ge n(s_0) $, since $ m = 3 $ and $ n = 4 $, rendering it unstable. 
(b) Size distributions $P(S)$ of unstable clusters generated by perturbations in various Net-I models.
    }
    \label{fig:cluster_pdf}
\end{figure}

\begin{figure}[tb]
    \centering
    \includegraphics[width= 8.6 cm]{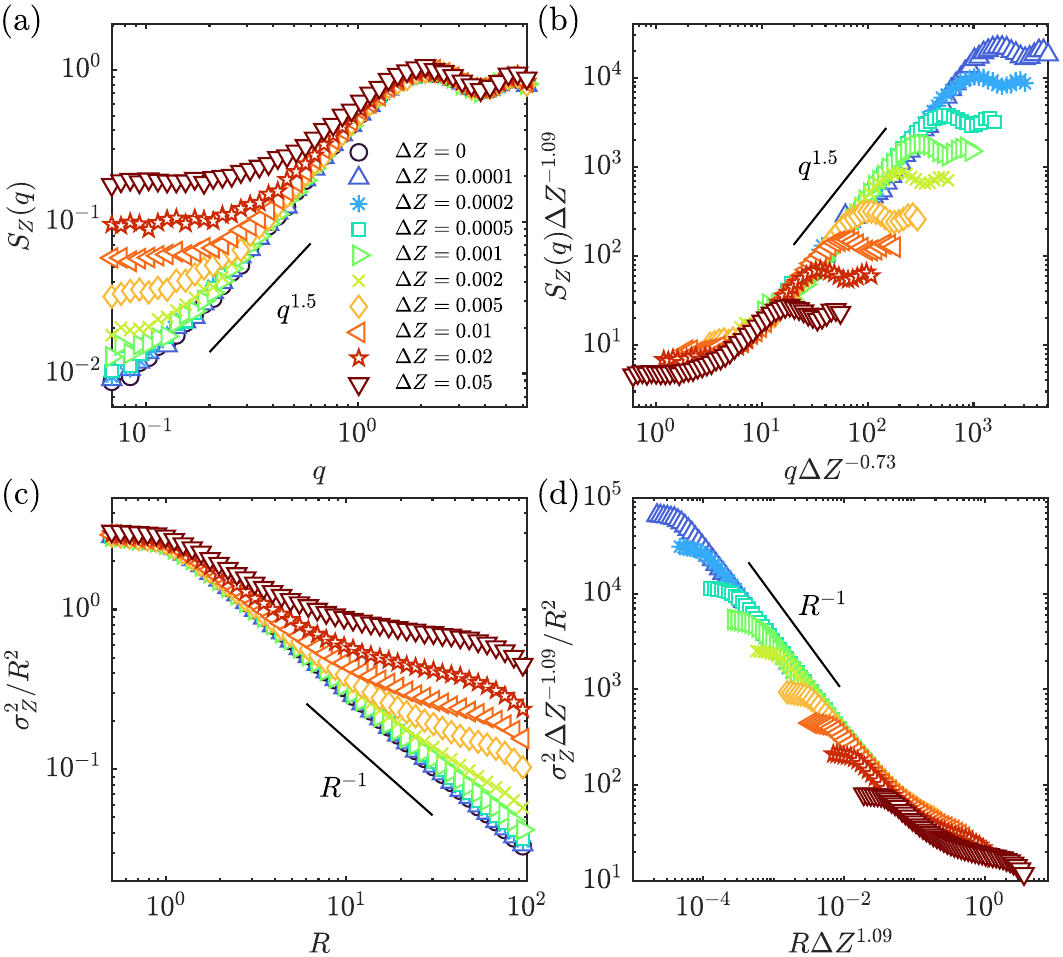}
    \caption{
    {\bf Diverging length scales approaching the TST in Net-I.}
    Contact-number fluctuations characterized using (a) the structure factor $S_Z(q)$ in the Fourier space and (c) the variance $\sigma_Z^2(R)$ in the real space, at different $\Delta Z$  on a 2D triangular lattice. 
    (b,d) The collapse of the rescaled data for networks close to the TST.}
    \label{fig:lengthscale}
\end{figure}

{\bf Introduction.}
Hyperuniformity, with widespread observations in classical \cite{Hexner2015PRL, Kurita2011PRE, Oguz2017PRB, Lei-Ni, Wilken2020PRL, Wilken2021PRL, Maher2023PRE, Hitin2024PRE}, quantum \cite{Florescu2009PNAS, Sakai-quantum}, biological \cite{HuangMingji2021PNAS, LiuYuan2024PRL}, mathematical \cite{Torquato2008JSM, Klatt-math}, and cosmological systems \cite{Gabrielli-cosmology}, describes the suppression of large-scale fluctuations in the spatial distribution of physical quantities \cite{Torquato2018PR, Torquato2016PRE}.
A physical quantity $A$ in $d$ dimensions is hyperuniform if its variance scales sub-extensively with observation radius $R$, 
\begin{equation}
\sigma_A^2 \propto R^{d-\gamma},
\end{equation}
with $\gamma \in (0,1]$. In Fourier space, this corresponds to a structure factor 
\begin{equation}
S_A(q) \propto q^\alpha,
\label{eq:S}
\end{equation}
as $q \to 0$, with $\alpha > 0$. 
Three hyperuniform classes are defined according to the values of exponents - Class I: $\gamma=1$ and $\alpha >1$; Class II: $\gamma = \alpha = 1$ with logarithmic corrections; Class III: $\gamma = \alpha \in (0,1)$~\cite{Torquato2018PR}. Recently, interest in hyperuniformity has been expanded beyond crystals and quasicrystals to disordered materials, owing to their unique optical and mechanical properties \cite{Florescu2009PNAS, Newby2025PRE, WangYinqiao2025NC}.

The particles  in equilibrium fluids are distributed uniformly at large distances ($\alpha =0$) in order to have  a finite isothermal compressibility, while crystals are typically hyperuniform ($\gamma=1$) due to the ordered arrangement  of particles.
For packings of randomly jammed particles, it was conjectured that the distribution of particle positions is hyperuniform (density hyperuniform, or DH)~\cite{Torquato2003PRE, Torquato2018PR}, although this conjecture has been debated~\cite{Ozawa2017SPP, Ikeda2017PRE, WuYegang2015PRE, Atkinson2016PRE, Ikeda2015PRE}. 
More robust hyperuniformity is observed in the distribution of mechanical contacts between particles (contact-number hyperuniformity, or CNH) at the jamming transition, with $\gamma \approx 1$ and $\alpha \approx 1.5$~\cite{Hexner2018PRL, Hexner2019PRL, franz2020critical,   Rissone2021PRL, WangYinqiao2025NC}.
Accompanied with CNH is two distinct diverging lengths $\xi_Z$ and $\xi_f$ characterized by different critical exponents 
{$\nu_Z$ and $\nu_f$}~\cite{Hexner2018PRL, Hexner2019PRL}.  The diverging lengths, in contrast to a finite nucleation length  at typical first-order phase transitions, imply jamming as a critical point. However, the diverging fluctuations ($\gamma < 0$) in standard critical phenomena seem incompatible with hyperuniformity ($\gamma > 0$). Although it has been assumed that the hyperuniformity at jamming is related to its criticality~\cite{Atkinson2016PRE}, the paradox is still unsolved. 

Great efforts have been made to uncover the origin of hyperuniformity. 
In systems with long-range interactions, density fluctuations are suppressed by a conserved field~\cite{stillinger1968general, goldfriend2017screening, Hexner2017PRL}, which however does not show critical behavior. 
 Hyperuniformity and criticality emerge in a class of non-equilibrium phase transitions, such as the  conserved directed percolation, which belong to the Manna universality class (MUC) \cite{Hexner2015PRL, Hexner2017PRL,Wiese2024PRL,Henkel-noneq}.
Recent works~\cite{Wilken2021PRL, Wilken2023PRL} propose to connect DH at jamming~\cite{Torquato-2000, O'Hern2002PRL, Ohern-PRE-2003, Wyart-PRL-2012, Milz2013PRE, Torquato2003PRE, Donev2005PRL, Maher2023PRE} to a dynamical absorbing phase transition in the biased random organization  model~\cite{Milz2013PRE, Ness2020PRL, Wilken2021PRL, Wilken2023PRL} in the MUC.
However, the MUC exponents $\gamma_{\rm MUC} = \alpha_{\rm MUC} = 0.45$ in 2D and $\gamma_{\rm MUC} = \alpha_{\rm MUC } = 0.25$ in 3D (Class III) are clearly different from the CNH exponent $\gamma \approx 1$ and $\alpha \approx 1.5$  (Class I) observed in simulations~\cite{Hexner2018PRL,Hexner2019PRL}. Thus the  origin of CNH at the jamming transition remains elusive.

Here, we propose a mechanism for CNH, rooted in two mechanical constraints: a global equality due to isostaticity at jamming and a local inequality ensuring mechanical stability of any sub-systems. Together, they suppress contact-number fluctuations to scale with the surface area ($R^{d-1}$) rather than the volume ($R^d$) of an observation window, closely paralleling DH in crystals.
Based on this mechanism, we propose a minimum network model with a topological satisfiability transition, which exhibits simultaneously CNH and critical behavior as in the jamming transition. The CNH and critical exponents are in agreement in the minimum model and granular packings, suggesting a new universality class of non-equilibrium phase transitions.
Our results show that  CNH can be decoupled from DH, as it arises even with uniformly distributed particle positions.
\\

{\bf  An intuitive argument for the origin of contact number hyperuniformity.}
We first revisit the DH in crystals. As shown in Fig.\ref{fig:diagram}(a), the fluctuation $\sigma_n^2(R)$ of the particle number $n(R)$ within a window of a radius $R$ scales with the surface area $R^{d-1}$, not the volume $R^{d}$. Analogously, we show below that, at the jamming transition, contact-number fluctuations $\sigma_Z^2(R)$ scale as $R^{d-1}$. The derivation starts from the following two types of constraints imposed on the contact network at the jamming transition.

(i) A global equality due {to} the isostatic condition,
\begin{equation}
M =  d N,
\label{eq:isostatic}
\end{equation}
where $M$ is the total number of contacts (bonds) or constraints ($2M = N Z$ with $Z$ the coordination number) and $dN$ is the degrees of freedom. 
Corrections from global translations and finite bulk modulus are $O(1/N)$~\cite{goodrich2012finite} and negligible here.

(ii) A local inequality required by the mechanical stability for any sub-system\cite{ellenbro2015rigidity} with a radius $R$, 
\begin{equation}
m(R) >  d \, n(R), \,\, {\rm for \,\, any\,\,} R,
\label{eq:local_stability}
\end{equation}
where $m(R)$ is the number of contacts connected to the particles within the sub-system, which has a total particle count denoted as $n(R)$.

As can be seen in Fig.\ref{fig:diagram}(b), 
$m(R) = m_{\rm I}(R) + m_{\rm B}(R)$, where $m_{\rm I}(R)$ is the number of contacts inside and $m_{\rm B}(R)$ the number of contacts crossing the boundary. Applying Eq.~(\ref{eq:local_stability}) to the complementary set of $m(R)$ yields $\overline{m}(R) \equiv M - m_{\rm I}(R) > d[N-n(R)]$, 
which, combined with Eq.~(\ref{eq:local_stability}), gives  
\begin{equation}
d \, n(R) - m_{\rm B}(R) < m_{\rm I}(R) < d \, n(R).
\label{eq:m_I}
\end{equation} 
Equation~(\ref{eq:m_I}) shows that the bulk contact number is bounded from two sides, and its fluctuation depends on the surface area, $\delta m(R) \sim \delta m_{\rm I}(R) \sim m_{\rm B}(R) \sim R^{d-1}$. 

Following Refs.~\cite{Hexner2018PRL,Hexner2019PRL}, we define the local coordination number $z(R) = \sum_{i \in s(R)} Z_i$ for particles inside the sub-system $s(R)$. By definition, $z(R)=m(R) + m_{\rm I}(R)$ (note that the $m_{\rm I}$ contacts should be double counted). Following a similar strategy as discussed above for $m(R)$, it can be shown that,
$2 d\, n(R) - d\, n_{\rm E}(R) < z(R)  < 2d\,n(R) + d\,n_{\rm O}(R)$,               
where $n_{\rm E}$ (inside) and $n_{\rm O}$ (outside) denote the number of boundary particles forming the $m_{\rm B}$ boundary contacts (see Fig. \ref{fig:diagram} (b)). 
Thus the fluctuation of $z$ also scales with the surface area, $\delta z(R)  \sim n_{\rm E}(R) + n_{\rm O}(R) \sim R^{d-1}$.

The above analysis demonstrates that combining the local inequality constraints Eq.~(\ref{eq:local_stability}) and the corresponding global equality constraint Eq.~(\ref{eq:isostatic}) naturally results in a hyperuniform distribution. 
At the jamming transition, both constraints arise from mechanical reasons.
More generally, one can consider systems governed by (i) $A = cN + k$ and (ii) $a(R) > c\, n(R) + k$ (or $a(R) < c\, n(R) + k$) for any sub-systems, where $A$ is a physical quantity and $c, k$ are constants. Any system with such constraints - regardless of the underlying physics - should exhibit hyperuniformity. The key is that the complement of a sub-system must also satisfy the inequality; together with the global constraint, this bounds $a(R)$ on both sides, limiting its fluctuations to scale with the surface area.\\


{\bf A minimum network model with contact number hyperuniformity.}
To further reveal the origin of CNH, we introduce a minimal network model composed of  sites distributed  according to given geometric patterns. The sites are randomly {connected} to their neighbors under constraints analogous to Eqs.~(\ref{eq:isostatic}) and (\ref{eq:local_stability}). Identifying networks that satisfy these constraints poses a complex optimization problem; even verifying the satisfiability of a given configuration is with time complexity scaling as $\propto 2^N$. To maintain tractability, we focus on models solvable via polynomial-time algorithms.

The model, called Net-I,  satisfies global and local constraints:
\begin{equation}
 M = N - 1, 
\label{eq:NetI-global}
\end{equation}
and 
\begin{equation}
m(s) \geq n(s), \,\, {\rm for \,\,any \,\, } s,
\label{eq:NetI-local}
\end{equation}
where $s$ denotes an arbitrary sub-system.
It forms a spanning, acyclic tree with each node connected to exactly one neighbor, generated via an optimized Monte Carlo algorithm. We systematically explore various base networks in  2D, 3D and 4D (see Methods: Generation of Net-I). 
For comparison, we also analyze  packings of frictionless spheres generated via rapid quench and swap algorithms {(see Fig.~\ref{fig:diagram}(c), and Methods: Generation of frictionless sphere packings)}, representing generic and ultra-stable jammed states, respectively. 

Spatial fluctuations in contact numbers are quantified by the variance $\sigma_Z^2(R)$ within a circle of a radius $R$ (see Methods: contact number fluctuations). As shown in Fig.~\ref{fig:fluactuation}(a), $\sigma_Z^2(R) \propto R^{d-\gamma}$, with $\gamma \approx 1$ in multiple Net-I networks - a scaling also observed in rapidly quenched and ultrastable granular packings (Fig.~\ref{fig:fluactuation} (d)). 
This exponent $\gamma = 1$ holds across dimensions $d = 2$--4, in contrast to the MUC exponents: $\gamma_{\mathrm{MUC}} = 0.45$ in 2D and $0.25$ in 3D. Figure~\ref{fig:fluactuation}(a) indicates that CNH and DH are not necessarily correlated. In particular, random triangulation networks exhibit a uniform site distribution (the pair correlation function $g(r) = 1$), yet CNH can still arise
{(for a more rigorous formulation, see Methods: contact number fluctuations).}


We have argued that the contact number fluctuations inside a sub-system depend on the number of sites near the boundary of the sub-system (Eq.~\ref{eq:m_I}). Simulation data from network models and granular packings can directly verify this insight. Because $n_{\rm E}(R) \approx n_{\rm O}(R)$ for large windows, for the scaling analysis, we focus on the $\sigma_Z^2(R)$ vs. $n_{\rm E}(R)$ data (the $\sigma_Z^2(R)$ vs $n_{\rm O}(R)$ data give the same conclusion). Figure~\ref{fig:fluactuation} (b,e) confirm a linear relationship in the large-$R$ limit, 
\begin{equation}
\sigma_Z^2 (R) = X \, n_{\rm E} (R).
\label{eq:X}
\end{equation}
The pre-factor $X$ is obtained by linear fitting: $X=0.49(5)$ for Net-I; $X = 1.01(1)$ and $1.6(3)$ for granular packings in 2D and 3D respectively. 
The errors are given by the standard deviation between the fitting for different curves plus the range of the 95\% confidence interval.
{Interestingly, we observe a universal relation, $X\approx c/2$, where $c$ is the prefactor in the constraints ($c=1$ for Net-I and $c=d$ for granular packings).} 

 Assuming contact changes (connection or disconnection events) are independent events with a probability $p=1/2$, it can be shown that {$X=c/2$} 
 (see Methods: Derivation of  $X=c/2$), which is consistent with our numerical results. In more general cases (see Net-II in SI), contact changes are dependent and {$X \neq c/2$;}
 the value of {$\hat{X} \equiv 2 X/c$} then characterizes  more delicate correlations between contacts beyond hyperuniformity.

Hyperuniformity can also be characterized in Fourier space by the structure factor of the contact number, $S_Z(q)$ (defined in Methods: Structure factor of the contact number), through its scaling behavior
Eq.~(\ref{eq:S})~\cite{Torquato2003PRE, Torquato2018PR,Hexner2018PRL}. For Class I hyperuniform systems ($\gamma=1$ and $\alpha > 1$), the precise value of $\alpha$ can reveal subtle structural features.
In Net-I  and granular packings (Fig.~\ref{fig:fluactuation} (c) and (f)), $\alpha \approx 1.5$ is universally observed, independent of the dimensionality, the site patterns in Net-I,  and the preparation protocol of packings. 
From fitting, we obtain $\alpha=1.51(7)$ in 2D and $\alpha = 1.48(8) $ in 3D {for Net-I}. 
{The robustness of hyperuniformity with respect to packing preparation protocols is illustrated in Extended Data Fig.~\ref{fig:3dpackings}.}
Our results imply a potential correspondence between $\alpha$ and $\hat{X}$:  both of them might be related to subtle spatial fluctuations that cannot be captured by the hyperuniformity scaling $\sigma_Z^2 \propto R^{d-1}$. \\

\begin{table*}[!htbp]
\centering
\caption{{\bf Exponents for the TST in Net-I,  granular packings, and the Manna universality class.} 
The exponent $\alpha$ in the Manna class is estimated using the value of $\gamma$  based on the relation $\alpha = \gamma$ for Class III hyperuniformity. The Manna class does not have exponents corresponding to $\nu_Z$ and $\nu_f$; for comparison, we list the exponents $\nu_\perp$ and  $\nu_\parallel$ associated to the divergence of two correlation lengths $\xi_\perp$ and $\xi_\parallel$.
}
\begin{tabular}{ c | c c  c  | c c c |c }
\hline
   & \multicolumn{3}{c|}{$d=2$} & \multicolumn{3}{c|}{$d=3$} & expectation\\\
     & TST & packing &  Manna & TST & packing & Manna  & TST \\
\hline
$\gamma$ & 1 & 1 &  0.45(3)~\cite{Hexner2015PRL} & 1 & 1 & 0.24(7)~\cite{Hexner2015PRL} & 1 \\
$\alpha$  & 1.51(7)  & 1.53(4)~\cite{Hexner2018PRL}  & 0.45 &  1.48(8) & 1.52(5)~\cite{Hexner2018PRL} & 0.24 & 3/2 \\
$\tau$ & 1.35(1) & 1.34-1.36~\cite{Lin2014PNAS} & {1.39(2)~\cite{Henkel-noneq}} & 1.45(1) & 1.45-1.48~\cite{Lin2014PNAS} & {1.46(3)~\cite{Henkel-noneq}} & 3/2 \\
$\nu_Z$ & 0.73(9) & 0.7(1)~\cite{Hexner2018PRL} & - &  0.65(9) & 0.85(15)~\cite{Hexner2018PRL} & - & 2/3 \\

$\nu_f$ & 1.09(13) & 1.07(18)~\cite{Hexner2018PRL} & - & 0.98(14) & 1.29(27)~\cite{Hexner2018PRL} & - & 1\\
$\nu_\perp$ & - & - & 0.799(14)~\cite{Henkel-noneq} & - & - & 0.593(13)~\cite{Henkel-noneq} 
& - \\
$\nu_\parallel$ & - & - & 1.225(29)~\cite{Henkel-noneq} & - & - & 1.081(27)~\cite{Henkel-noneq} & - \\
\hline 
\hline
\end{tabular}
\label{tab:exponents}
\end{table*}

{\bf Critical Scalings in the minimum model near a topological satisfiability transition.}
In the above, we have demonstrated that when the two constraints are satisfied, the Net-I displays CNH. If the total number of contacts $M$ is treated as a variable, then it is easy to see that the condition $M=N-1$ corresponds to  a {\it topological satisfiability transition} (TST). 
(i) When $M<M_{\rm c} \equiv N-1$, the number of contacts is fewer than the number of sites, and 
it is impossible to satisfy all local  constraints Eq.~(\ref{eq:NetI-local}).
 (ii) When $M = M_{\rm c}$, Eq.~(\ref{eq:NetI-local}) is satisfied marginally  and the system has strict CNH. Meanwhile, we will show below that the system is also critical, as revealed by a power-law distribution of unstable clusters with one contact swapped. (iii) When $M > M_{\rm c}$, the network has redundant sites to satisfy Eq.~(\ref{eq:NetI-local}),
and it can be shown that the system has CNH up to a finite length scale.

The above picture has an analogue in frictionless granular packings near the jamming transition. (i) When $Z< Z_{\rm iso} \equiv 2d$, the unjammed system is unstable, meaning that
the local stability condition Eq.~(\ref{eq:local_stability}) cannot be satisfied for arbitrary sub-systems.
 (ii) At the jamming transition $Z=Z_{\rm iso}$, the system has CNH as shown in Ref.~\cite{Hexner2018PRL, Hexner2019PRL}. Meanwhile, isostaticity manifests that the system is in a marginally stable state \cite{Wyart2005PRE, Muller2015ARCMP}, i.e., it becomes unstable upon removing one contact or applying an infinitesimal perturbation. 
 In this case, perturbations, such as shear deformations, typically induce avalanche behavior, leading to power law distributions of physical quantities such as the stress drop. (iii) Away from the jamming transition, $Z> Z_{\rm iso}$, the system is hyperstatic, i.e, Eq.~(\ref{eq:local_stability}) is satisfied with redundant contacts. 
Ref.~\cite{Hexner2018PRL} shows that the deviation from the perfect CNH is captured by two distinct length scales $\xi_Z$ and $\xi_f$, correspondingly associated  with  the two-point correlations of the local contact numbers and the contact-number fluctuations $\sigma_Z^2(R)$.
The exponents in the scalings $\xi_Z \propto \Delta Z^{-\nu_Z}$ and $\xi_f \propto \Delta Z^{-\nu_f}$ are numerically determined in~\cite{Hexner2018PRL} by collapsing  $S_Z(q)$ data following a scaling function  $S_Z(q)=\Delta Z^{\beta} f(q \xi_Z)$ for different $\Delta Z$ and using the relation $\nu_f = \beta = \alpha  \nu_Z$
(see Table~\ref{tab:exponents}).

In the following, we will first show the criticality in Net-I at the TST ($M=N-1$). We apply a perturbation by randomly breaking a connection and then randomly connecting an unconnected  pair of sites. Applying the perturbation induces sub-systems that violate the given constraint $m(s) \geq n(s)$, as shown in Fig.~\ref{fig:cluster_pdf} (a). An unstable cluster is defined as the connected component with $m(s)<n(s)$, or equivalently the minimum sub-system violating the constraint $m(s) \geq n(s)$, for a given random realization of perturbation. 
The size $S$ of the unstable cluster is defined by the number of sites within it. Adding a single connection to the unstable cluster can make the entire system satisfying all constraints. 
Fig.~\ref{fig:cluster_pdf} (b) shows the probability distributions $P(S)$ across various dimensions and site patterns, exhibiting power-law behavior, $P(S)\propto S^{-\tau}$, with an exponent    $\tau^{2\mathrm{D}} = 1.35(1)$, $\tau^{3\mathrm{D}} = 1.45(1)$, and  $\tau^{4\mathrm{D}} = 1.49(2)$. 
Assuming $P(S)$ is analogous to the distribution of avalanche sizes under perturbations, we can compare our values to those reported in the literature. 
Mean-field theories have been largely developed to describe   a class of phenomena called ``cracking noise'', which generally give $\tau^{\rm MF} = 3/2$~\cite{sethna1993hysteresis,Sethna2001Nature}.
Based on a numerical elastoplastic model and theoretical scaling relations, Lin, et al. obtain respectively $\tau^{\rm{2D-packing}} = 1.36$ and $1.34$ in  2D, and $\tau^{\rm{3D-packing}} = 1.45$ and $1.48$ in  3D~\cite{Lin2014PNAS,Lin2015PRL}. The mean-field glass theory with a full-replica symmetry breaking ansatz  predicts $\tau_{\rm J}\approx1.42$ in large dimensions~\cite{Franz2017PRE}.
Although the exact value of $\tau$ remains indecisive, the  values measured in Net-I are nevertheless close to those of packings reported in the literature.

Next we analyze the diverging length scales near the TST.
Following Ref.~\cite{Hexner2018PRL}, we compute  $S_Z(q)$ at different $\Delta Z\equiv Z-Z_{\mathrm{c}}$, where $Z \equiv 2M/N$ and $Z_{\rm c} \equiv 2M_{\rm c}/N =  2(N-1)/N$, and collapse the data according to the scaling form   $S_Z(q)=\Delta Z^{\beta} f(q \xi_Z)$ (see Fig.~\ref{fig:lengthscale} (a,b)), 
where $f(x)$ is a constant for $x \ll 1$, and $f(x) \propto x^{\alpha}$ for $x \gg 1$. To recover the small-$q$ behavior  $S_Z(q) \sim q^\alpha$, the relation $\beta =\alpha \nu_Z$ is required~\cite{Hexner2018PRL}.
The best collapse yields $\nu_Z^{\mathrm{2D}}=0.73(9)$ and $\beta^{\mathrm{2D}} = 1.09(13)$ (see Fig.~\ref{fig:lengthscale} (b)).

Taking into account the asymptotic relationship between the fluctuation and the structure factor \cite{Torquato2018PR,Hexner2018PRL}, $\lim_{R\to\infty}\sigma_Z^2(R)/N(R) = \lim_{q\to0}S_Z(q)\propto\Delta Z^{\beta}$, we can express the fluctuation as $\sigma_Z^2 (R)/R^d = \Delta Z^{\beta} g(R\xi_f)$, where $g(x)$ is constant for $x\gg1$ and $g(x)\propto x^{-1}$ for $x \ll 1$.
On the other hand, $\lim_{R\to\infty}\sigma_Z^2(R)/R^d \sim (R/\xi_f)^{-1}$.
Matching $\Delta Z^{\beta}$ with $(R/\xi_f)^{-1}$ implies a relation $\nu_f = \beta$, i.e., $\nu_f^{\mathrm{2D}} = 1.09(13)$.
Figures~\ref{fig:lengthscale} (c, d) show that the $\sigma_Z^2(R)$ curves and their excellent collapse after rescaling using $\nu_f^{\mathrm{2D}} =\beta = 1.09$, except for deviations at large scales due to finite-size effects. 
The obtained exponents,  $\nu_Z^{\mathrm{2D}}=0.73(9)$ and $\nu_f^{\mathrm{2D}} = 1.09(13)$,  are independent of site patterns {(see Extended Data Fig.~\ref{fig:NetI2D4})}, and  are consistent with the reported values for 2D packings, $\nu_Z^{\rm{2D-packing}} = 0.7(1)$ and  
$\nu_f^{\rm{2D-packing}} = 1.07(18)$~\cite{Hexner2018PRL}.

In 3D, the fitted exponents  are $\nu_Z^{\mathrm{3D}} = 0.65(9)$ and $\nu_f^{\mathrm{3D}} = 0.98(14)$ in Net-I {(see Extended Data Fig.~\ref{fig:NetI3D6})}, which are slightly different from 
$\nu_Z^{\rm{3D-packing}} = 0.85(15)$ and $\nu_f^{\rm{3D-packing}} = 1.29 (27)$ in 3D packings~\cite{Hexner2018PRL}. 
We emphasize that it is very challenging to measure accurate  exponents in 3D, due to strong finite-size effects in data collapsing~\cite{Hexner2018PRL}.

Combining our numerical results with several scaling arguments, we provide below theoretical conjectures of these exponents (see Table~\ref{tab:exponents}).   Both Net-I and packing data suggest that $\gamma =1$ and $\alpha = 3/2$, independent of $d$. The exponent $\tau$ is close to the mean-field value $\tau=3/2$, with weak $d$-dependence on the tail of $P(S)$ (Fig.~\ref{fig:cluster_pdf}b). To derive the exponent $\nu_f$, we apply a scaling argument similar to the ``cutting argument'' by Wyart, et al.~\cite{Wyart2005EPL}. 
When $\Delta Z>0$, the entire system does not satisfy the global  constraint Eq.~(\ref{eq:NetI-global}), but one can ask up to what length scale, the sub-system can satisfy Eq.~(\ref{eq:NetI-global}). 
In a sub-system of a linear size $\xi_f$,
the increase of contact numbers in the bulk $\Delta Z \xi_f^d$ should be compensated by the surface effect $Z \xi_f^{d-1}$, yielding $\Delta Z \xi_f^d \sim Z \xi_f^{d-1}$, 
 or $\xi_f \sim \Delta Z^{-1}$ ($\nu_f=1$).
Thus $\xi_f$ is a crossover length: when $R<\xi_f$, the contact distribution $\sigma_Z^2(R) \sim R^{d-1}$ is hyperuniform; when $R \gg \xi_f$, it becomes  uniform  $\sigma_Z^2(R) \sim R^{d}$. Finally, using the relation $\nu_f = \alpha \nu_Z$, we obtain $\nu_Z = 2/3$.
As compared in Table~\ref{tab:exponents}, the theoretical expectations agree reasonably well with numerical values considering  numerical errors, in both Net-I and packings. Moreover, the exponents do not seem to {be} dependent on $d$, implying that both Net-I and packings are mean-field-like, consistent with the previous  proposal of an upper critical dimension  $d_{\rm u}=2$~\cite{goodrich2012finite}. On the other hand, it is clear that these exponents do not belong to the Manna universality class.\\

{\bf Discussion.}
We have presented a simple and intuitive physical picture of CNH at the jamming transition of frictionless spheres: it arises due to the combined effects of global and local constraints required by isostaticity and mechanical stability. The minimum network model (Net-I) reproduces the  exponents for CNH and critical scalings near the jamming transition, suggesting jamming as a TST. Our network models are mathematically well-defined and could be analytically tractable in future studies.

It is important to discuss the difference between the TST in Net-I and the SAT-UNSAT transition in random constraint satisfaction problems (rCSPs) including  $k$-satisfiability and $q$-coloring problems~\cite{krzakala2007gibbs}. The SAT-UNSAT transition in rCSPs has been considered sharing the universality classes of hard sphere jamming~\cite{franz2017universality, yoshino2018disorder}, based on the following analogue: the satisfiable (SAT) phase corresponds to the unjammed phase where hard spheres can be positioned without overlapping, and the unsatisfiable (UNSAT) phase corresponds to the over-jammed phase where overlapping cannot be avoided (i.e., hard sphere solutions do not exist).  In this sense, the SAT-UNSAT transition in rCSPs can be referred to as a ``geometrical'' satisfiability transition. In contrast, in this study we consider a ``topological'' satisfiability transition, where the constraints are imposed on the contact network topology instead of particle positions. In our case, the over-jammed phase is SAT and the unjammed phase is UNSAT. Our results show that {CNH} is originated at topological constraints, which can be decoupled from geometrical constraints~\cite{franz2020critical, xing2024origin}. 

The TST is different from the rigidity percolation (RP)~\cite{jacobs1995generic,  ellenbro2015rigidity}.
The RP refers to the emergence of a system-spanning rigid cluster - the other smaller clusters do not have to be rigid. The largest cluster has a heterogeneous, fractal shape, typical for  a second-order phase transition. In contrast, at the TST, not only the largest cluster, but also any sub-system is required to satisfy the local stability constraint. In Net-I, the system is percolated at the TST ($M = M_{\rm c}$), but does not have to be percolated either below or above. In other words, the TST does not correspond to a percolation transition. 
The essential difference between the TST and RP perhaps explains the dramatic differences in the nature of the marginal states at the jamming transition and the RP~\cite{ellenbro2015rigidity}.

Our study opens the door to several future directions.
(i) Experimental verification. 
It remains unclear if CNH would emerge in experimental  packings, since frictional or non-spherical particles can induce uncertain constraints and exhibit strong dependence on preparation protocols \cite{Wyart2005EPL,vanHecke2010JPCM}.
In addition, boundary conditions can significantly affect the results~\cite{Hexner2019PRL}, requiring careful treatment for non-periodic systems. 
(ii) Dynamics. To capture the mechanical and rhelogical properties in the unjammed phase, one needs to extend the current static network models to dynamic ones. 
(iii) Extension to other systems. It is very interesting to seek  other systems, e.g., cosmological models~\cite{Gabrielli-cosmology}, with similar global and local constraints, and examine the universality of the discovered hyeruniformity~\cite{Gabrielli-cosmology}.
\\

{\bf Acknowledgments.} This work is supported by the NSFC (No. 11974238 and No. 12274291). This work is also supported by the Innovation Program of Shanghai Municipal Education Commission under No. 2021-01-07-00-02-E00138. We also acknowledge the support from the Student Innovation Center of Shanghai Jiao Tong University. Y.J. acknowledges support from NSFC (Grants 11974361, 12161141007, 11935002, and 12047503), from Chinese Academy of Sciences (Grants ZDBS-LY-7017 and KGFZD-145-22-13), and from Wenzhou Institute (Grant WIUCASICTP2022). D.P. acknowledges support from NSFC (Grants 12404290).

\bibliography{references}

\clearpage
\centerline{\Large \bf Methods}
\hspace{1cm}

{\bf Generation of Net-I.}
We begin by selecting a base network structure and randomly generating an initial configuration of $M$ connections. This configuration may not satisfy the local inequality conditions required by all sub-systems. An optimization algorithm is then used to iteratively adjust connection positions, progressively improving the satisfaction of these conditions.
When $M = N - 1$, the local conditions require the network to form a spanning tree. To expedite convergence, we prioritize breaking connections within cycles and reconnecting disjoint components.
For $M \geq N$, each connected component $s_c$ must satisfy
$m(s_c) \geq n(s_c)$,
where $m(s_c)$ and $n(s_c)$ denote the number of connections and sites in component $s_c$, respectively. Consequently, the algorithm removes excess connections from components with $m(s_c) > n(s_c)$ and adds connections to those with $m(s_c) < n(s_c)$.
To reduce bias from the initialization process, once all local conditions are met, the algorithm continues to adjust connections while preserving the constraints, thoroughly exploring the configuration space.

We systematically investigate the following site patterns: in 2D, (1) square lattices, (2) triangular lattices with nearest neighbor connections (NNCs), (3) Union Jack lattice (square lattice with diagonal next NNCs), and (4) random triangulation network (Delaunay triangulation of a Poisson point process); in 3D, (5) cubic lattice and (6) face-centered cubic (FCC) lattice with NNCs; in 4D, (7) hypercubic lattice with NNCs.\\

{\bf Generation of frictionless sphere packings.}
We generate random configurations with a packing density $\varphi = 0.70$ that is far above the  jamming density. 
Then, the systems are decompressed, under quasistatic ahtermal conditions,  to 
a state very close to the jamming transition ($\delta \varphi \approx 10^{-5}$). For over-jammed systems, we employ harmonic interaction and minimize the total potential energy. Packings of frictionless particles with different size dispersities are prepared.
Three representative size distributions are generated {in 3D}: (1) monodisperse, (2) binary distribution of a diameter ratio 1.4 with equal number of large and small particles, (3) polydisperse with an inverse power-law distribution, $P(r) \sim r^{-3}$  with a size ratio $r_{\rm max}/r_{\rm min} = 0.45$. In 2D,  bidisperse configurations are generated.

Jammed states with a varying stability  are prepared by quenching the thermally equilibrated configurations~\cite{pan2023_pnas_n,pan2022nonlinea}. Well equilibrated samples with $\varphi_{\rm eq} = 0.598,~0.609,~0.619,~0.630$ are generated via the SWAP algorithm~\cite{ninarell2017models}. 
We focus on systems with inverse power-law size distribution, due to its high efficiency~\cite{ninarell2017models}. 
The systems are composed by $N = 192000$ and $N = 16000$ particles for random and stable configurations, respectively.

When calculating the contact number fluctuations $\sigma_Z^2(R)$ and structure factor $S_Z(q)$, all rattlers are excluded, and the length scale is normalized by the average distance between contacting particles.\\

{\bf Contact number fluctuations.}
We define the variance of contact numbers within a spherical window as
\begin{equation}
    \sigma_Z^2(R) = \left \langle \left ( \sum_i \delta Z_i w\left ( \mathbf{r}_i-\mathbf{x}_0; R \right ) \right )^2 \right \rangle,
\end{equation}
where $\delta Z_i=Z_i-Z$ denotes the deviation of the contact number $Z_i$ at a site $i$ from the average $Z$, $\mathbf{r}_i$ represents the spatial coordinate of the site, and $w$ represents a spherical window function of radius $R$ centered at a randomly sampled point $\mathbf{x}_0$. The ensemble average $\langle \cdot \rangle$ is taken for all $\mathbf{x}_0$ of a given network and for different random network realizations.

With this definition, CNH and DH can be independent.
To show that, consider the contact number correlation function,
\begin{equation}
    C_Z(r)=\left \langle \frac{\sum_{i,j}\delta (|\mathbf{r}_i-\mathbf{r}_j|-r)\delta Z_i \delta Z_j}{\sum_{i,j}\delta (|\mathbf{r}_i-\mathbf{r}_j|-r)} \right \rangle,
\end{equation}
where $\delta(r)$ is a radial Dirac delta function (we assume isotropic systems for convenience).
The contact number fluctuations can be expressed as
\begin{equation}
    \sigma_Z^2(R) = \rho V(R)\left [\left\langle\delta Z_i^2\right\rangle +\rho \int \mathrm{d}\mathbf{r}\ C_Z(r) g(r) \alpha_2(\mathbf{r};R) \right ].
\end{equation}
Here, $\rho$ denotes the mean density of the site number, $\left\langle\delta Z_i^2\right\rangle$ is the single-site fluctuation, 
and $\alpha_2$ quantifies the overlap of windows, $\alpha_2(\mathbf{r};R)=\int \mathrm{d}\mathbf{x}_0 \ w(\mathbf{x}_0;R)w(\mathbf{x}_0+\mathbf{r};R)/V(R)$. 
When the two-point correlation function $g(r)=1$, i.e., the absence of DH, as in random triangulation networks, $\sigma_Z^2(R)$ is controlled by $C_{Z}(r)$ -- in this case, CNH is obviously independent of the site distribution. \\

{\bf Derivation of  $X=c/2$.}
As mentioned in the main text, contact number fluctuations in sub-systems exhibit linear scaling with the number of boundary sites,
Eq.~(\ref{eq:X}).
The pre-factor $X$ reflects the space correlations in contact numbers to a certain extent.
To see that, we start with the simplest case, Net-I.
For Net-I, the limitation of $z(R)$ is imposed by $2n(R)-n_{\rm E}(R)\le z(R) \le 2n(R)+n_{\rm O}(R)$. 
We assume that each time the contact number within a sub-system changes by one with an equal probability $p=1/2$ and the $n_{\rm B}=2n_{\rm E}$ events are created as independent events. 
Then the distribution $p(Z)$ is a binomial distribution 
\begin{equation}
p(z) =  \frac{n_{\rm B}!}{n_{\rm B}!(n_{\rm B}-z)!}p^z(1-p)^{n-z}, 
\end{equation}
whose variance is $\sigma_Z^2 = n_{\rm B} p (1-p) = n_{\rm E}/2$ (i.e., $X=1/2$), consistent with the numerical result $X=0.49(5)$ for Net-I in Fig.~\ref{fig:fluactuation} (b).
Generalizing the above analysis to granular packings, where $n_{\rm B}=2d\,n_{\rm E}$ due to $2 d\, n(R) - d\, n_{\rm E}(R) < z(R)  < 2d\,n(R) + d\,n_{\rm O}(R)$ for packings, one expects $\sigma_Z^2 = d n_{\rm E}/2$ (i.e., $X=d/2$), consistent with the numerical results, $X=1.01(1)$ in 2D $X = 1.6(3)$ in 3D (Fig.~\ref{fig:fluactuation} (e)). \\

{\bf Structure factor of the contact number.}
We define the structure factor of the contact number $S_Z(q)$ as
\begin{equation}
    S_Z(q) = \frac{1}{N}\left \langle \sum_{i} \delta Z_i \mathrm{e}^{-\mathrm{i} \mathbf{q}\cdot \mathbf{r}_i} \right \rangle,
\end{equation}
where $\langle \cdot \rangle$ refers to the average over all samples and all directions of the wave vector $\mathbf{q}$. 
For hyperuniform systems, the structure factor approaches zero in the long-wavelength limit.

\onecolumngrid

\setcounter{figure}{0}
\renewcommand{\figurename}{Extended Data Fig.}

\begin{figure*}[p!]
    \centering
    \includegraphics[width= 17.8 cm]{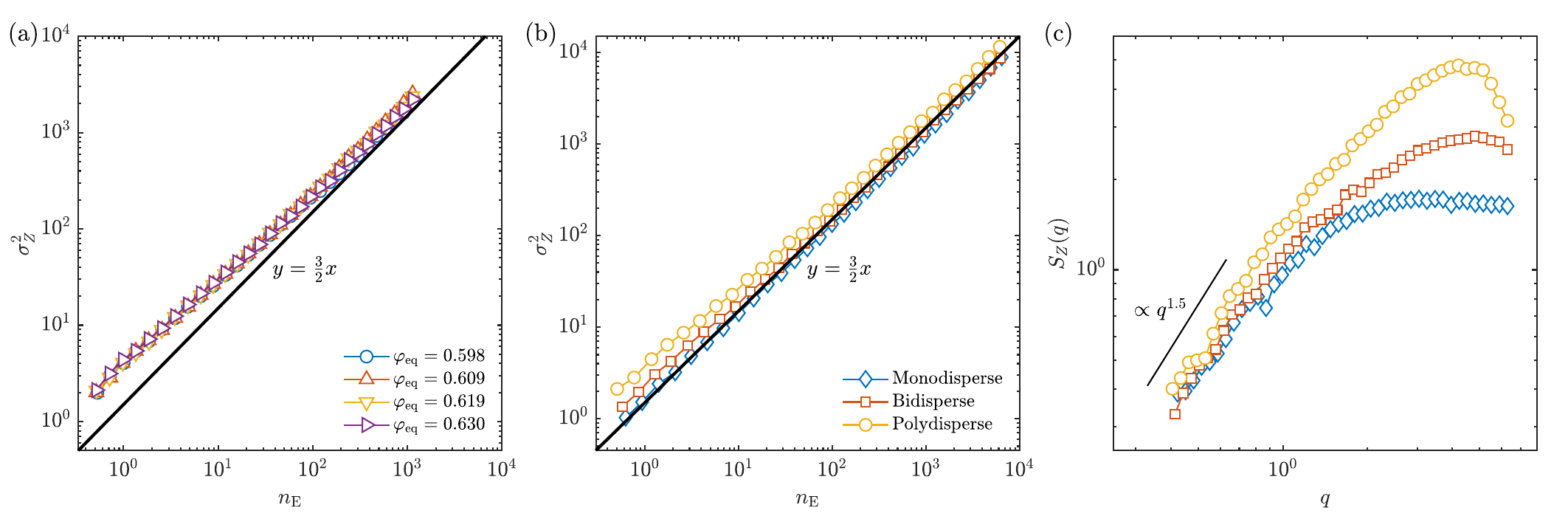}
    \caption{
    {\bf CNH in 3D sphere packings obtained by different preparation protocols.} (a) The contact number variance $\sigma^2_Z$ versus the boundary particle number $n_{\rm E}$ for polydisperse frictionless sphere packings prepared by quenching from initially equilibrium configurations at different  $\varphi_{\rm eq}$. 
    (b) $\sigma^2_Z$ versus $n_{\rm E}$ for packings with different size dispersities. 
    The dispersity of particles affects contact number fluctuations only at small scales.
   (c) The contact-number structure factor $S_Z(q)$ for packings with different size dispersities. The $S_Z(q)$ data curves exhibit clear separation at large $q$, but approach zero with the same scaling as $q\to 0$.
    }
    \label{fig:3dpackings}
\end{figure*}

\newpage
\begin{figure*}[t!]
    \centering
    \includegraphics[width= 17.8 cm]{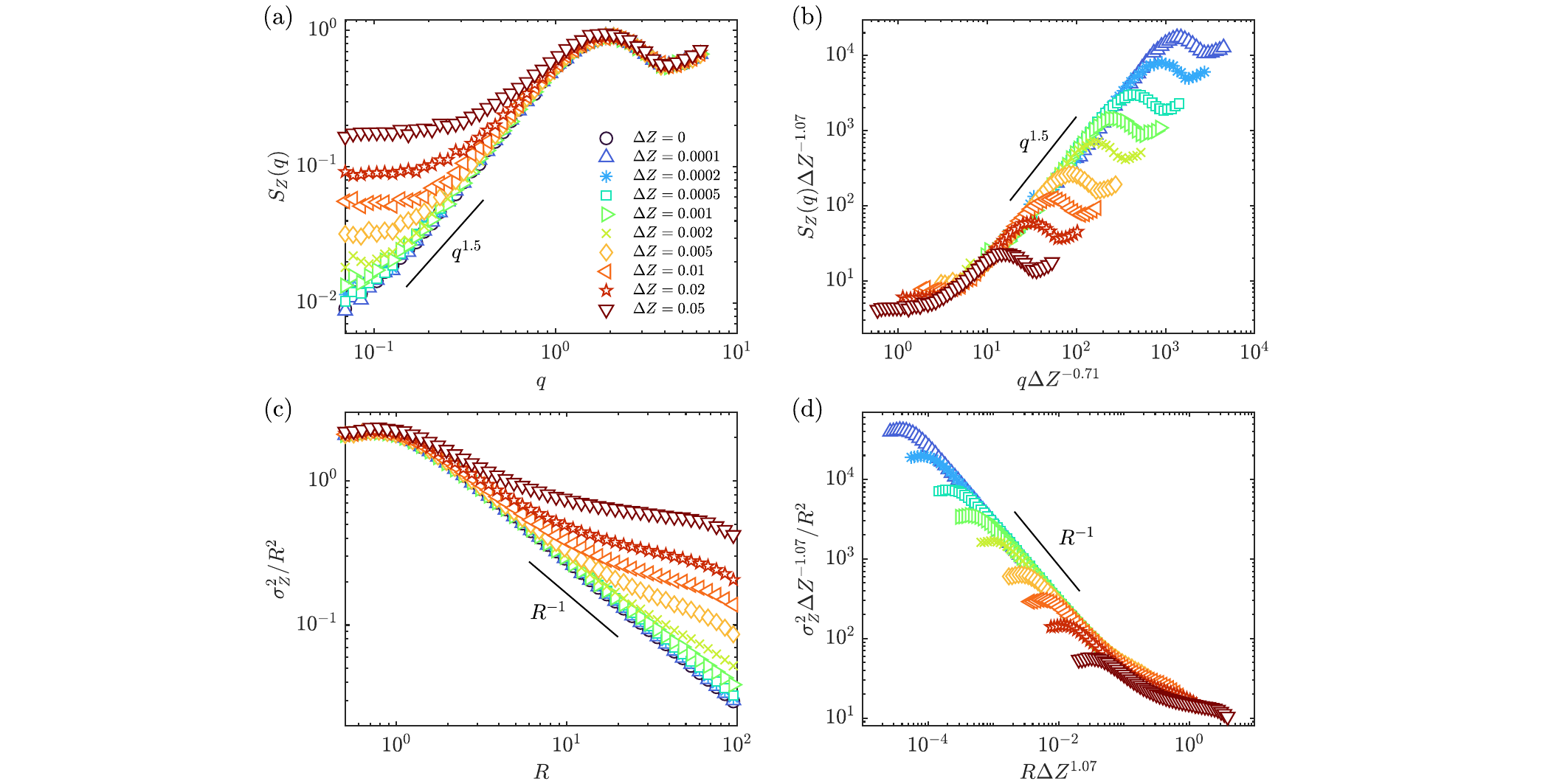}
    \caption{{\bf Diverging length scales approaching the TST in Net-I on 2D square lattices.}
    (a) Contact-number structure factor $S_Z(q)$ 
and (c) variance $\sigma_Z^2(R)$. The best collapse fitting indicates that two length scales diverge with critical exponents $\nu_Z^{\mathrm{2D}}=0.71(10)$ and $\nu_f^{\mathrm{2D}}=1.07(15)$, respectively. The collapses of the rescaled data are shown in (b) and (d).
    }
    \label{fig:NetI2D4}
\end{figure*}

\newpage
\begin{figure*}[t!]
    \centering
    \includegraphics[width= 17.8 cm]{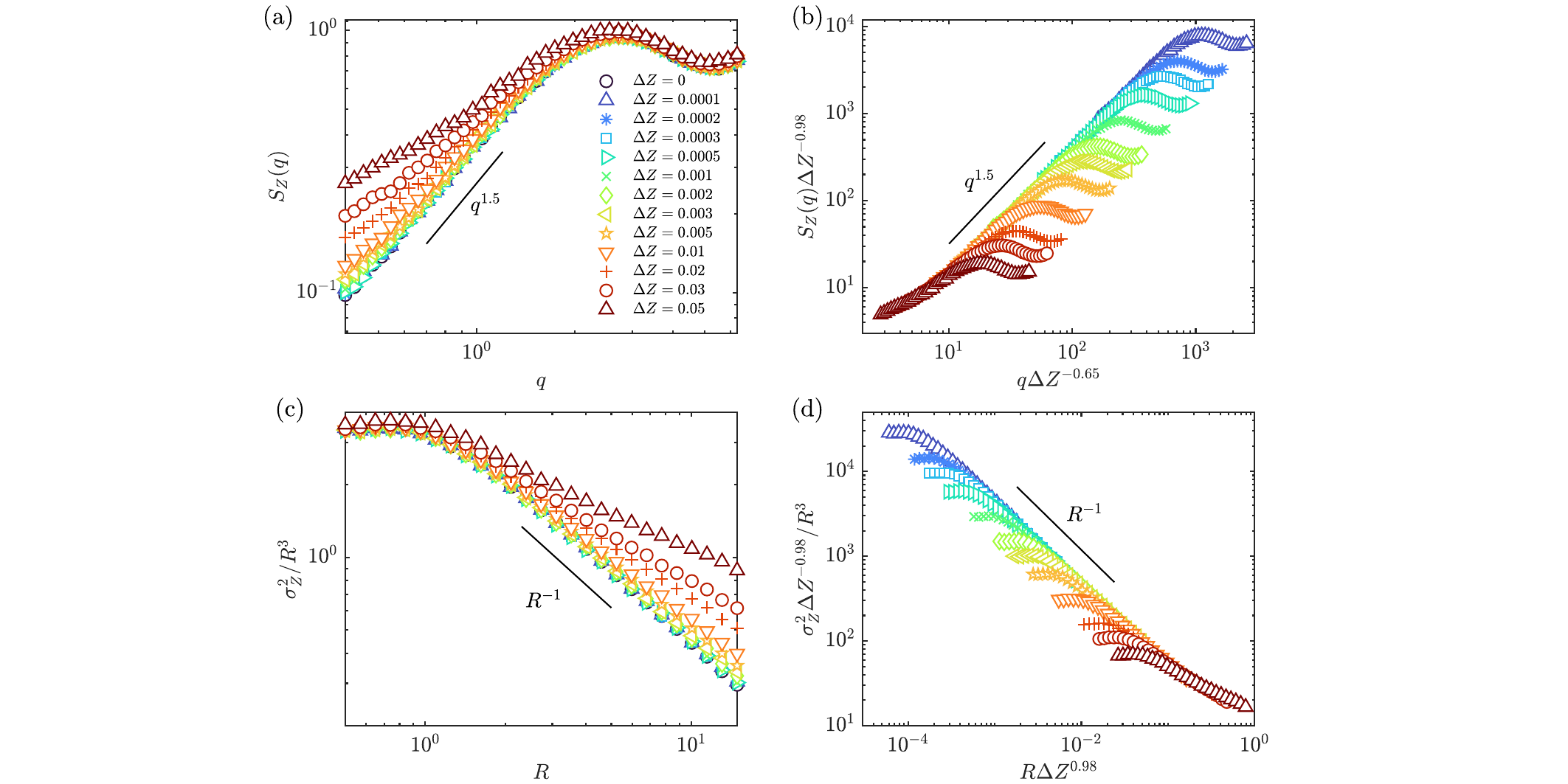}
    \caption{{\bf Diverging length scales approaching the TST in Net-I on 3D cubic lattices.}
    (a) Contact-number structure factor $S_Z(q)$ and (c) variance $\sigma_Z^2(R)$. The best collapse fitting indicates that two length scales diverge with critical exponents $\nu_Z^{\mathrm{3D}}=0.65(9)$ and $\nu_f^{\mathrm{3D}}=0.98(14)$, respectively. The collapses of the rescaled data are shown in (b) and (d).
    }
    \label{fig:NetI3D6}
\end{figure*}

\clearpage

\centerline{\bf \Large  Supplementary Material}

\setcounter{figure}{0}
\setcounter{equation}{0}
\setcounter{table}{0}
\renewcommand{\figurename}{Fig.}
\renewcommand\thefigure{S\arabic{figure}}
\renewcommand\theequation{S\arabic{equation}}
\renewcommand\thesection{S\arabic{section}}
\renewcommand\thetable{S\arabic{table}}


\section*{An extended  network model with contact number hyperuniformity}

\begin{figure*}
    \centering
    \includegraphics[width= 17.8 cm]{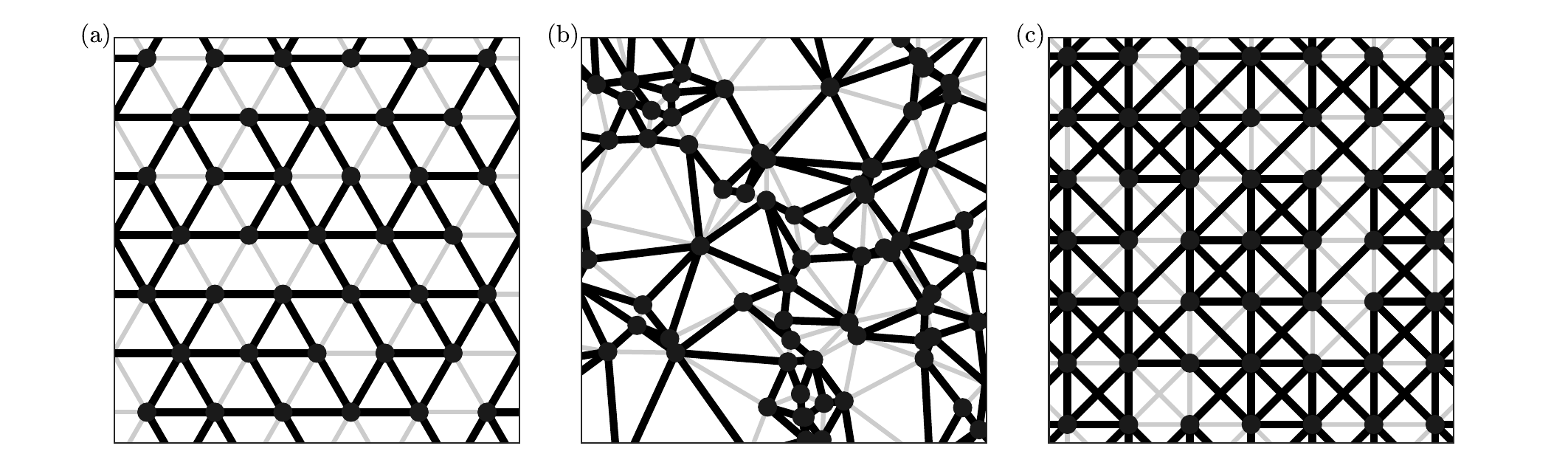}
    \caption{{\bf Schematic diagrams of Net-II models.} (a) A local configuration of Net-II on the triangular lattice with $c=2$.  (b) Net-II on a random triangulated network with $c=2$. (c) Net-II on a Union Jack lattice network with $c=3$. 
    }
    \label{fig:diagramNet-II}
\end{figure*}

The Net-I model, although simple, captures many quantitative features of CNH in granular packings near the jamming transition. The pre-factor $d$ in Eqs.~(\ref{eq:isostatic}) and~(\ref{eq:local_stability}) is replaced by a constant (one) in any dimensions. Naturally, we can ask if 
it is possible to construct network models with a flexible pre-factor. For this purpose, we design an extended model, called Net-II, satisfying (i) $M=c(N-1)+1$ globally and (ii) $m(s)\ge cn(s)+1$ locally, where $c$ is a positive integer parameter.
{Several examples of Net-II configurations are presented in Fig.~\ref{fig:diagramNet-II}}. Note that the sites in Net-II are still distributed on given base patterns. Constraints on the arrangement of particle positions in packings are not included in Net-II.  For example, local stability requires that the contact points on a particle  cannot be on the same semi-sphere for pure repulsive interactions --  such geometrical constraints are not considered in our model.

Net-II  configurations  cannot be generated by simple Monte Carlo methods anymore. Instead, we find these configurations by mapping the problem to solving a set of homogeneous equations.
Constraints are met if no solution variable vanishes, reducing constraint verification to solving homogeneous equations efficiently via singular value decomposition.
For Net-II, we focus on 2D systems due to computational efficiency limitations. Diagrams of selected 2D networks are shown in Fig.~\ref{fig:diagramNet-II}.

We begin with a base site pattern and randomly create $M$ connections. 
For the given network, analogous to force balance equations, we map all connections to $M$ variables,  constructing a homogeneous linear equation system. We denote  the variable corresponding to the connection between sites $i$ and $j$ by $x_{ij}=x_{ji}>0$, and  an unconnected pair by $x_{ij}= 0$. 
Each node corresponds to $c$ homogeneous equations:
\begin{equation}
    \sum_{j=1}^N a_{ij}^k x_{ij} =0, \ \ k= 1,2,...,c,\ \ i=1,2,...,N,
\end{equation}
where $a_{ij}^k$ are random coefficients (e.g., a uniform distribution in $[-\pi/2, \pi/2]$) satisfying $a_{ij}^k = -a_{ji}^k$. This constructs a homogeneous system with $M$ independent variables and $cN$ equations. Under periodic boundary conditions, the maximal rank of the coefficient matrix is $c(N-1)$. For $M=c(N-1)+1$, the null space dimensionality $\ge 1$. When constraints are unsatisfied, certain variables are always zero, corresponding to regions lacking connections. 
We employ singular value decomposition to obtain normalized nontrivial solutions.  We set a threshold $T=10^{-13}$ and repeat with three different $\{a^k_{ij}\}$. The variables falling below $T$ are regarded  as zero. 
Employing an optimization algorithm, at each step we randomly reposition a connection, until the number of zero variables is progressively decreased to zero. 
Thereafter, we continue to explore the configuration space without violating constraints.

\begin{figure*}
    \centering
    \includegraphics[width= 20cm]{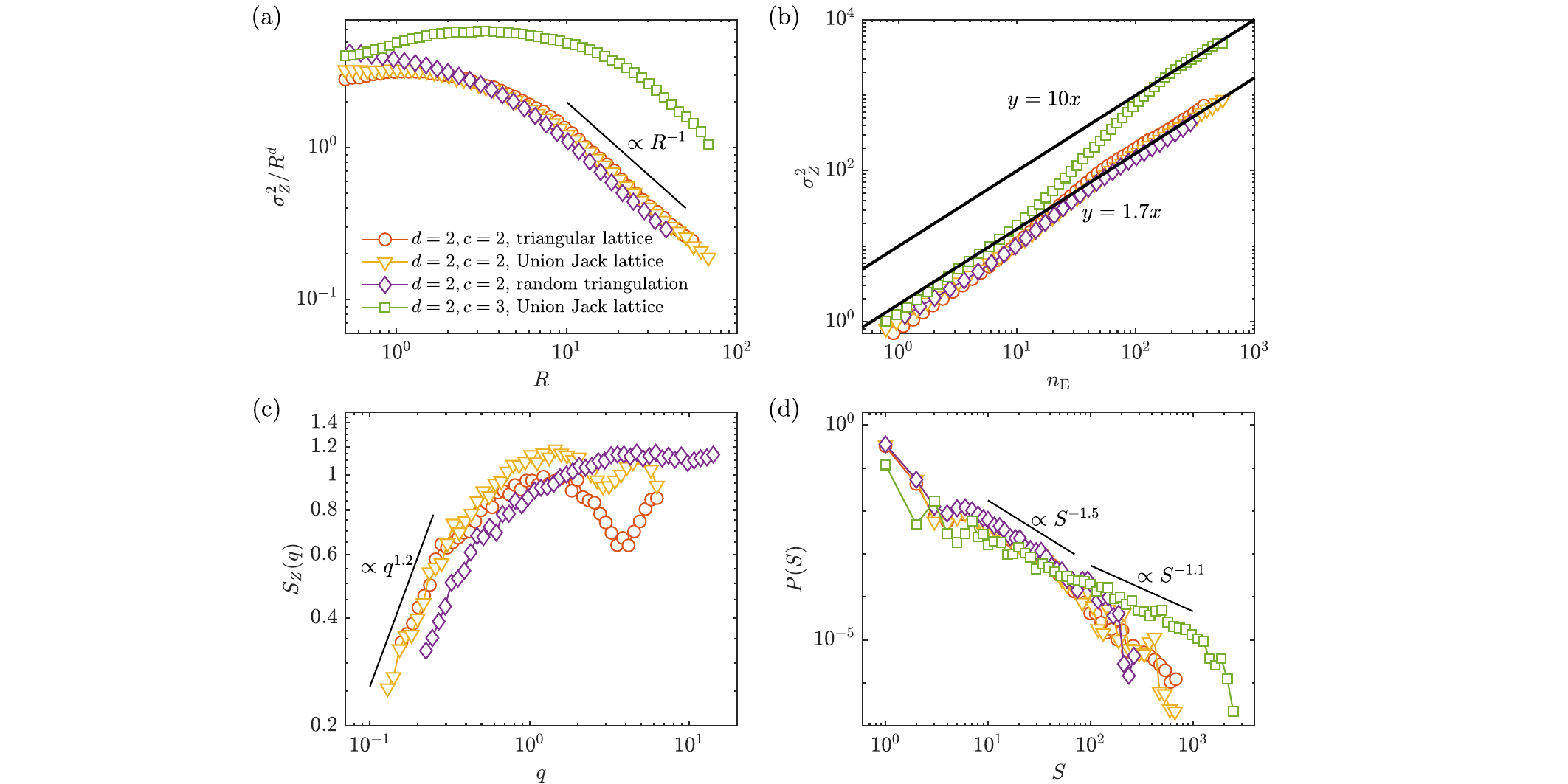}
    \caption{
    {\bf CNH and unstable cluster size distributions in Net-II.}  (a) The contact number variance $\sigma^2_Z$ versus $R$. (b) The dependence of $\sigma^2_Z$ on the number of boundary sites $n_{\rm E}$. 
    Linear fitting gives $X= 1.7(3) $ and $10.0(3)$ for $c=2$ and $3$ respectively. 
    (c) The contact-number structure factor $S_Z(q)$. 
    The exponent $\alpha = 1.2(2)$ for $c=2$.
    When $c=3$, $\alpha$ is undetermined  due to insufficient system sizes.
    (d) Unstable cluster size distribution $P(S)$ after a connection swap perturbation. The distributions exhibit a power-law scaling, although finite-size effects limit the measurable range. The exponent $\tau$ obtained from fitting: for $c=2$, $\tau = 1.5(2)$, while for $c=3$, $\tau = 1.07(5)$. 
    }
    \label{fig:Net-II}
\end{figure*}

As shown by Fig. \ref{fig:Net-II}, the Net-II also exhibits CNH, characterized by 
$\sigma_Z^2(q) \propto  R^{d-1}$ at large $R$ and $S(q) \propto q^\alpha$ at small $q$. However, several scaling exponents and parameters have different values now. 
In Net-II with $c=2$,  we find the exponent $\alpha = 1.2(2)$ {(Fig. \ref{fig:Net-II})}, which is different from $\alpha \approx 1.5$ in Net-I and granular packings. The prefactor $X$ in Eq.~(\ref{eq:X}) is $X= 1.7(3) $ and $10.0(3)$ with $c=2$ and $3$ respectively, which do not obey the relation $X=c/2$. 
Fitting the power-law   $P(S) \sim S^{-\tau}$ gives $\tau=1.5(2)$ for $c=2$ and $\tau=1.07(5)$ for $c=3$, although we cannot exclude the possibility of a universal, $d$-independent  $\tau$ in both Net-I and Net-II.
In short, Net-II reproduces key features of CNH, but it appears that subtle correlations, characterized by $\alpha$ and $X$, are enhanced compared to Net-I.

\end{document}